\documentclass[runningheads]{llncs}
\usepackage{graphicx}
\usepackage{tabularx}
\usepackage{multirow}
\usepackage{float}
\usepackage{amssymb}
\usepackage{amsmath}
\usepackage{hyperref}
\usepackage{tablefootnote}
\usepackage[dvipsnames]{xcolor}
\usepackage{makecell}
\usepackage{multirow}
\usepackage{bm}
\usepackage{booktabs, makecell}
\usepackage{tabu}

\newcommand{\mcc}[1]{\multicolumn{1}{c}{#1}} 

\begin{document}
\title{Domain Adaptation of Echocardiography Segmentation Via Reinforcement Learning}
\titlerunning{RL4Seg: Domain Adaptation for Segmentation via Reinforcement Learning}
%
%


\author{Arnaud Judge\inst{1} \and
Thierry Judge \inst{1,2} \and
Nicolas Duchateau \inst{2,3} \and
Roman A. Sandler \inst{4} \and
Joseph Z. Sokol \inst{4} \and
Olivier Bernard\inst{2} \and
Pierre-Marc Jodoin \inst{1}}

\authorrunning{A. Judge et al.}


\institute{Department of Computer Science, University of Sherbrooke, Sherbrooke, QC, Canada \and
University of Lyon, CREATIS, CNRS UMR5220, Inserm U1294, INSA-Lyon 
\and 
Institut Universitaire de France (IUF) \and 
iCardio.ai}



%
\maketitle              
\begin{abstract}

Performance of deep learning segmentation models is significantly challenged in its transferability across different medical imaging domains, 
particularly when aiming to adapt these models to a target domain with insufficient annotated data for effective fine-tuning. While existing domain adaptation (DA) methods propose strategies to alleviate this problem, these methods do not explicitly incorporate human-verified segmentation priors, compromising the potential of a model to produce anatomically plausible segmentations. We introduce RL4Seg, an innovative reinforcement learning framework that reduces the need to otherwise incorporate large expertly annotated datasets in the target domain, and eliminates the need for lengthy manual human review. Using a target dataset of 10,000 unannotated 2D echocardiographic images, RL4Seg not only outperforms existing state-of-the-art DA methods in accuracy but also achieves 99\% anatomical validity on a subset of 220 expert-validated subjects from the target domain. Furthermore, our framework's reward network offers uncertainty estimates comparable with dedicated state-of-the-art uncertainty methods, demonstrating the utility and effectiveness of RL4Seg in overcoming domain adaptation challenges in medical image segmentation.





%

\end{abstract}
\section{Introduction}
\vspace{-0.1cm}

Image segmentation using deep neural networks is accurate and reliable on many medical applications, including 2D echocardiography \cite{camus,Chen:FCM:2020}. However, knowledge acquired from one domain (e.g. high quality segmentations on one dataset) does not confer to easy transferability to another without fine-tuning. To this end, domain adaptation (DA) aims to bridge the gap between datasets by limiting (sometimes removing) the amount of annotations required on a new dataset \cite{Guan:TBE:2022}. Leveraging unlabeled data is essential for DA methods as the collection of such data is inexpensive compared to the time needed for their labeling. 


Many methods use pseudo-labels to learn from unlabeled data on the target domain \cite{Li-pseudolabels-review}. Pseudo-labels are obtained from the predictions of a pre-trained model on the target domain and can be used either by a second model dedicated to the target domain \cite{ReciprocalLearning} or by fine-tuning the same source model
\cite{UDAS}. However, this may introduce inaccurate information into the training process. Confidence based pseudo-labeling was recently introduced to handle this issue \cite{TS-IT,shen2023cotraining}. These methods integrate an additional sub-network to assess the quality of the generated pseudo-labels and weight their influence during training. Other DA techniques include image-to-image translation \cite{iacono2023cycleGAN}, namely matching the distributions of images from the source to the target domain before training. However, these methods do not explicitly consider anatomical correctness, resulting in segmentations with reasonable Dice but poor anatomical validity.

Reinforcement learning (RL) is widely used for a variety of tasks requiring an intelligent agent. Notably, RL from human feedback (RLHF) is used in language processing to obtain outputs aligned with human preferences \cite{ziegler2019fine,NEURIPS2020_RLSummarize,ouyang2022_instructGPT}. ChatGPT is a popular example of this methodology's success. However, applications of RL to image segmentation remain mostly limited to accessory tasks such as hyper-parameter tuning or region of interest detection \cite{hu2023reinforcement}.


In this paper, we propose {\em RL4Seg}, a novel DA framework orthogonal to all previous works. 
The framework uses RL to bridge the gap between source and target domains and ensure high rates of anatomical validity of the target segmentations on a large dataset of 10,000 unannotated images. Taking inspiration from ChatGPT's protocol in learning how to output text aligned with human preferences, our model learns to output segmentations aligned with anatomical validity metrics, eliminating the need for expert interaction and annotations thus making the method self-supervised. In addition to image segmentation, our framework optimizes an error prediction network which is shown to be competitive with standalone state-of-the-art uncertainty methods\footnote{\scriptsize Code is available at https://github.com/arnaudjudge/RL4Seg}.

\vspace{-0.1cm}
\section{Method}
\vspace{-0.1cm}

Considering a target dataset $\mathcal{D}_T = \{x_T^{(1)}...\:x_T^{(n)}\}$ containing only images and a fully annotated source dataset $\mathcal{D}_S = \{(x_S^{(1)}, y_S^{(1)}) ...(x_S^{(m)}, y_S^{(m)})\}$, our method uses RL to optimize a neural network for segmenting images from a target domain. We illustrate the efficiency of our method on 2D echocardiographic images, one of the most challenging modalities for segmentation. Please note that our framework is generic and can fit other modalities and applications.

\begin{figure}[t]
\includegraphics[width=\textwidth]{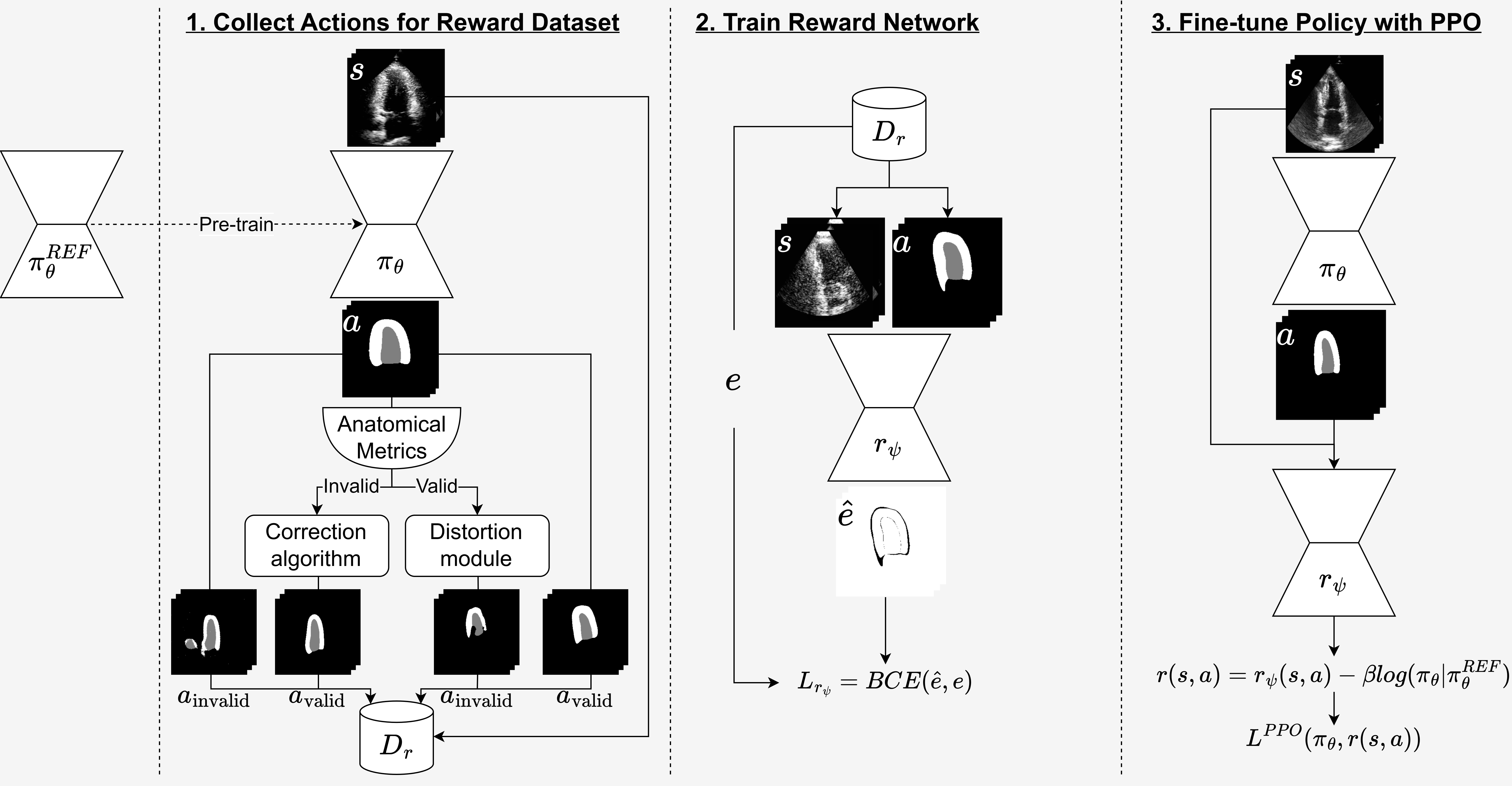}
\caption{RL4Seg, divided in three sections: [left] the reward dataset creation, [mid] the reward network training, and [right] the fine-tuning of the policy. \vspace{-0.4cm}} 
\label{fig:method}
\end{figure}

\vspace{-0.3cm}
\subsection{Reinforcement Learning}



In the typical RL context, problems are posed as trajectories of states and actions following a Markov Decision Process \cite{Sutton:Book:2018}. At each time step $t$ in the trajectory, a reward $r(s_t, a_t)$ is obtained for the action $a_t$ taken by a policy $\pi$ at the current state $s_t$ and a new state $s_{t+1}$ is reached. The policy is optimized with regards to the reward in order to maximize expected returns along a trajectory.

In the RL formalism, the value function $V^\pi$ corresponds to the expected total reward the current policy $\pi$ shall cumulate from state $s_t$ until the end of the trajectory. 
It considers all possible actions that could be taken at state $s_t$ by $\pi$, and can be expressed through Bellman's value function: 
\begin{align}
\label{bellman_value}
    V^\pi(s_t) &= \underset{s_{t+1} \sim P(\cdot|s_t, a_t)}{\underset{a_t \sim \pi(\cdot| s_t)}{\mathbb{E}}}[r(s_t, a_t) + \gamma V^\pi(s_{t+1})],
\end{align}
where $\gamma \in [0, 1]$ lowers the reward of actions further in time. Another key element is the Q function, which represents the total current and future reward for taking action $a_t$ at state $s_t$, considering the current policy $\pi$. Bellman's equation for this function 
is the expectation under state transition probabilities of the current reward, plus the discounted Q values for all subsequent state-action pairs:
\begin{align}
\label{bellman_q}
    Q^\pi(s_t, a_t) &= \underset{s_{t+1} \sim P(\cdot|s_t, a_t)}{\mathbb{E}}[r(s_t, a_t) + \gamma Q^\pi(s_{t+1}, a_{t+1})].
\end{align}

Finally, an advantage function is defined as:
$A(s_t, a_t) = Q^\pi(s_t, a_t) - V^\pi(s_t)$.
It describes the quality of the action taken given all possible actions the policy could take at that state, assuming the same policy dictates all future actions.

\vspace{-0.3cm}
\subsection{Segmentation RL}
Through the lens of RL, the image segmentation problem involves trajectories of length one. The state ($s=s_0$) is the input image, while the action ($a=a_0$) corresponds to the predicted output of a segmentation model, i.e. the policy $\pi$. The reward $r(s,a)$ is the pixel-wise accuracy of the segmentation map $a$ (the action) of a given image $s$ (the state).


\paragraph{The Policy} $\pi_\theta: \mathbb{R}^{H \times W}\!\!\! \rightarrow\!\! [0, 1]^{K \times H \times W}$ is a neural network (U-Net) with parameters $\theta$  that outputs an action $a$ (segmentation) given an input state $s$ (image). $K$ is the number of classes, and $H\times W$ is the image size.  $\pi_\theta$ outputs a probability distribution over all possible actions (segmentations) via a \textit{Softmax} function. This distribution is categorical over each pixel. During training, the actions are sampled from this distribution to explore the action space.

\paragraph{The Reward} $r_\psi: \mathbb{R}^{2 \times H \times W} \!\!\!\rightarrow \!\! [0, 1]^{H \times W}$ is a second neural network (U-Net) with parameters $\psi$ that estimates the reward for a given state $s$ and action $a$ (an image/segmentation pair). The reward is a pixel-wise error map of the given segmentation $a$. $r_\psi$ has a \textit{Sigmoid} output activation function. It is trained on the reward dataset $\mathcal{D}_r$ (Sec.~\ref{sec:reward_ds}), with a binary cross-entropy (BCE) loss function. 

\paragraph{The Q, Value and Advantage Functions:}
in this single timestep context, the Q and value functions do not contain any future state-action pairs and state transition probabilities: 
$V^\pi(s) = \underset{a \sim \pi(\cdot | s)}{\mathbb{E}} [r(s, a)]$, and $Q^\pi(s, a) = r(s, a)$.
The Q function approximates the reward exactly while the value function approximates the expected reward at state $s$ under the current policy. The advantage 
becomes $A(s, a) = r(s, a) - V^\pi(s)$,
which 
estimates the quality of the segmentation action $a$ compared to the average segmentation action the policy can take. 

\paragraph{The Value operator} $V_{\phi}^{\pi}: \mathbb{R}^{H \times W} \rightarrow [0, 1]^{H \times W}$ is a third neural network (U-Net) with parameters $\phi$ that approximates the value function. Its input is the state $s$ (the image), and its output is the anticipated reward map given the policy $\pi_{\theta}$. It has a \textit{Sigmoid} output activation function, as the possible rewards are in $[0,1]$. \\ 

Given the Reward and the Value networks $r_\psi$ and $V^{\pi}_\phi$, the advantage is computed by subtracting their predictions : $A(s,a) = r_\psi - V^{\pi}_{\phi}$.

\vspace{-0.3cm}
\subsection{RL4Seg}
\vspace{-0.2cm}

In the spirit of ChatGPT, our RL framework consists of three steps (Fig~\ref{fig:method}):
\begin{enumerate}
    \addtocounter{enumi}{-1}
    \item (Initialization) Before starting the RL loop, the segmentation neural network $\pi_\theta^{REF}$ is pre-trained on the fully annotated source dataset $\mathcal{D}_S$.  $\pi_\theta^{REF}$ will stand as the first version of the target policy $\pi_\theta$.  
    \item The policy $\pi_\theta$ first segments a subset of $N$ images from the unannotated target dataset $\mathcal{D}_T$. Following a procedure described below, these segmentation maps are then post-processed and stored in a reward dataset $\mathcal{D}_r$.
    \item The reward network $r_\psi(s,a)$ is trained on the reward dataset $\mathcal{D}_r$ to predict the error map $e$ of a segmentation mask $a$ associated to an image $s$.
    \item A copy of the policy is stored in $\pi^{old}_\theta$. Then, $\pi_\theta$ and $V_\phi^\pi$ are optimized with the newly trained reward model $r_\psi$ and the target dataset $\mathcal{D}_T$ using the PPO RL algorithm (see below).
\end{enumerate}
Steps 1 to 3 are repeated, each time with new samples from the unannotated target data to improve the policy $\pi_\theta$ and the reward network $r_\psi$. This goes on until every image of the target dataset $\mathcal{D}_T$ has been segmented.


\vspace{-0.2cm}
\subsubsection{Reward Dataset $\mathcal{D}_r$}
\label{sec:reward_ds}
It consists of pairs of images and segmentation masks $(s^{i},a^{i})$ as well as their corresponding error map $e^{i}$.  $\mathcal{D}_r$ is used to train the reward network $r_\psi$, which aims to predict $e^{i}$ given $(s^{i},a^{i})$.

First empty, $\mathcal{D}_r$ is populated during step 1 of the RL procedure. At each iteration, the policy $\pi_\theta$ segments a subset of $N$ images from the target dataset, resulting in correct segmentation maps for some images and incorrect segmentations for others. Since these images are unannotated, the anatomical validity of the segmentation maps is determined based on prior knowledge about the segmented organ. In our case, we assess the validity of the segmentation maps with 10 cardiac anatomical metrics inspired by \cite{Painchaud_metrics} (cf. Supplementary Material). 

Each segmentation map $a^i$ containing an anatomical error is post-processed with a dedicated warping system~\cite{Painchaud_2022}. This system implements a variational autoencoder (VAE) that warps an anatomically invalid shape towards its closest valid shape (see \cite{Painchaud_2022} for more details). The post-processed mask $\hat a^i$, its associated image $s^i$, the invalid mask $a^i$ and the pixel-wise difference $e^i$ between the corrected and invalid masks 
are then stored in the reward dataset $\mathcal{D}_r$.

As for the anatomically valid segmentation maps $a^i$, many pairs of valid and invalid segmentations are created and added to $\mathcal{D}_r$, using different perturbations independently applied to the policy weights $\theta$, the input image, and segmentations, 
aiming to simulate possible errors that a policy $\pi_\theta$ could produce.

\vspace{-0.2cm}
\subsubsection{Proximal Policy Optimization (PPO)}
Optimization of the policy $\pi_\theta$ is done according to the actor-critic style PPO algorithm \cite{schulman2017proximal}.  In this paper, a two-term loss function is used :  $L^{PPO}=L^{CLIP}+\alpha L^H$. 


$L^{CLIP}(\theta) = \mathbb{E}_\theta [\textrm{min}(\rho(\theta)A,\ \textrm{clip}(\rho(\theta), 1 - \epsilon, 1 + \epsilon)A)]$ is the {\em clipped surrogate loss} of the advantage function $A$ pursuing two objectives.
First, with the ratio $\rho(\theta) = \frac{\pi_\theta(a|s)}{\pi_{\theta}^{old}(a|s)}$\footnote{\scriptsize $\pi_\theta(a|s)$ represents the probability of taking action $a$ given state $s$, under the policy $\pi_\theta$}, it favors an increase of the output probability of the policy $\pi_\theta$ compared to the old policy $\pi_\theta^{old}$ for high-reward segmentations and decreases probabilities for low-reward segmentations.  Second, clipping $\rho(\theta)$ between $[1-\epsilon, 1+\epsilon]$ ensures that the policy updates remain reasonably small (we use $\epsilon=0.2$), whereas the $min$ operator allows for larger optimization steps in the direction of higher advantage when a previous update has led the policy to output actions with a worse outcome. The second loss term is $L_H = -\sum \pi_\theta \log(\pi_\theta )$, an entropy penalty on the policy's output distribution to ensure sufficient exploration. 


The reward $r(s,a)$ used to calculate the advantage 
is obtained with the reward network $r_\psi$ and a logarithmic penalty term to prevent the current policy $\pi_\theta$ from diverging from the reference $\pi_\theta^{REF}$ (the policy trained on $\mathcal{D}_S$):
\begin{align} 
    r(s, a) &= r_\psi(s, a) - \beta (\textup{log}\pi_\theta(a|s) - \textup{log}\pi_\theta^{REF}(a|s)).
\end{align}
where $\beta = 0.05$ in our experiments.
When creating the reward dataset $\mathcal{D}_r$
, the anatomically valid actions $a^i$ are kept as gold standards for PPO. They are substituted into PPO in place of the actions taken by the policy, and their reward is set to $1$ (maximum value) for all pixels. Thus, the PPO algorithm increases the probabilities that the policy will output such a segmentation map.

\subsubsection{Uncertainty Estimation}
Once trained, the reward network $r_\psi$ can serve as an uncertainty estimator by computing the complement to one of its output. In this way, high error probability areas have high uncertainty and vice-versa. For $r_\psi$ to output calibrated uncertainty maps, temperature scaling \cite{guo2017calibration} is applied during inference, using a scaling factor calculated with the validation set.

\begin{table}[tp]   
\caption{Results on the target data (average $\pm$ std. over 3 seeds) vs. the source dataset intra-expert variability. See Supplementary Material for an ablation study.}
    \centering
    \small
    \begin{tabular*}{\textwidth}{l @{\extracolsep{\fill}} ccccccc}
    \toprule
    \multirow{2}{*}{Method} & \multicolumn{3}{c}{Dice (\%) $\uparrow$} & \multicolumn{3}{c}{Hausdorff (mm)  $\downarrow$}   & \multirow{2}{*}{\makecell{Anatomical \\ Validity (\%)}$\uparrow$} \\
     \cmidrule(lr){2-4}  \cmidrule(lr){5-7}
    & \mcc{ENDO} &  \mcc{EPI} & \mcc{Avg.} & \mcc{ENDO} &  \mcc{EPI} & \mcc{Avg.}  & \\

\midrule\midrule

$\mathcal{D}_S$ intra-expert var.    & 94.4	&   95.4	&   94.9	&   4.3	&   5.0	&   4.6	&   100    \\
\midrule
Baseline (U-Net)       & 89.9\tiny{$\pm 0.2$}	&   93.7\tiny{$\pm 0.4$}	&   91.8\tiny{$\pm 0.2$}	&   7.0\tiny{$\pm 0.5$}	&   9.4\tiny{$\pm 1.0$}	&   8.2\tiny{$\pm 0.7$}	&   91.5\tiny{$\pm 1.4$}    \\
nnU-Net                 & 91.0\tiny{$\pm 0.1$}	&   94.6\tiny{$\pm 0.0$}	&   92.8\tiny{$\pm 0.0$}	&   6.3\tiny{$\pm 0.2$}	&   7.8\tiny{$\pm 0.4$}	&   7.1\tiny{$\pm 0.3$}	&   95.0\tiny{$\pm 0.7$}    \\
UDAS \cite{UDAS}       & 90.7\tiny{$\pm 0.3$}	&   93.7\tiny{$\pm 0.1$}	&   92.2\tiny{$\pm 0.2$}	&   6.7\tiny{$\pm 0.3$}	&   8.0\tiny{$\pm 0.5$}	&   7.3\tiny{$\pm 0.4$}	&   95.9\tiny{$\pm 1.0$}    \\
TS-IT \cite{TS-IT}     & 90.5\tiny{$\pm 0.1$}	&   93.6\tiny{$\pm 0.2$}	&   92.0\tiny{$\pm 0.2$}	&   6.1\tiny{$\pm 0.1$}	&   8.2\tiny{$\pm 0.4$}	&   7.1\tiny{$\pm 0.2$}	&   NA\tablefootnote{\scriptsize As the classes are segmented separately, anatomical validity cannot be computed reliably.}                      \\
RL4Seg (ours)          & \textbf{91.9\tiny{$\pm 0.1$}}	&   \textbf{94.7\tiny{$\pm 0.1$}}	&   \textbf{93.3\tiny{$\pm 0.0$}}	&   \textbf{4.9\tiny{$\pm 0.1$}}	&   \textbf{5.6\tiny{$\pm 0.1$}}	&   \textbf{5.3\tiny{$\pm 0.1$}}	&   \textbf{98.9\tiny{$\pm 0.8$} }   \\

    \bottomrule  & 
    \end{tabular*}
    
    \vspace{-0.6cm}
\label{tab:results}
\end{table}

\vspace{-0.2cm}
\section{Experiments}
\vspace{-0.1cm}

\noindent\textit{Source Dataset:}
500 echocardiography images (CAMUS dataset \cite{camus}) at end-diastole (ED) and end-systole (ES), in two- and four-chamber views, with left ventricle 
endocardium (ENDO) and epicardium (EPI) 
annotated by a cardiologist. The dataset was split into train-validation-test sets of 450-50-50 subjects. 
\noindent\textit{Target Dataset:} 10,000 unlabeled echocardiography images (at ED and ES) in two- and four-chamber views, from a heterogeneous private database, from various scanners and locations.
A subgroup of 220 subjects were annotated and manually validated by two experts to be used as the test set for all experiments.

\noindent\textit{Pre-processing and Post-processing:} All images in source and target datasets underwent identical preprocessing to bring the domains as close as possible. Images' contrast was increased locally through histogram equalization using \textit{scikit-image}'s \textit{exposure} package~\cite{scikit-image}.
All output segmentations were post-processed to remove any disconnected regions.


\noindent\textit{Model Configuration:} We used a U-Net with 7.8M parameters for all models and SOTA implementations (except nnU-Net). For 4 iterations ($N=2500$ target images) of our framework, training time was 5 hours with a NVIDIA 3090 GPU.

\begin{figure}[t] 
\includegraphics[width=\textwidth]{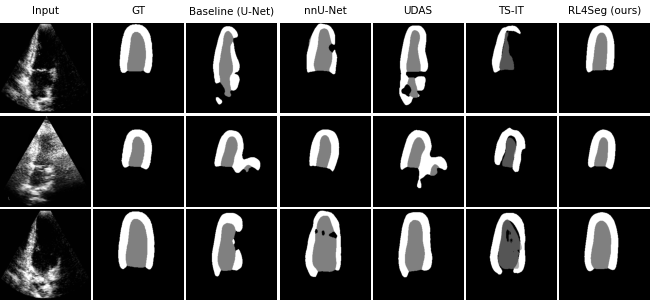}
\caption{Results for input images that the baseline was unable to segment correctly.} 
\label{fig:segmentation_examples}
\vspace{-0.5cm}
\end{figure}

\vspace{-0.25cm}
\subsubsection{Segmentation performance} 
We compared our framework (Tab.~\ref{tab:results}) with two segmentation methods: a U-Net and the nnU-Net\cite{Isensee2021-nnunet}, and two DA methods: {\em Self-Training of Early Features} (UDAS)\cite{UDAS} and {\em Transformation-Invariant Self-Training} (TS-IT)\cite{TS-IT}. UDAS trains the earliest layers of the network with pseudo-labels and a second segmentation head.  As for TS-IT, it uses confidence masked pseudo-labels. 
All models were tested on the same expert-validated test set from the target dataset. Metrics are Dice, Hausdorff distance and anatomical validity. 

All DA methods improve the segmentation results compared to simple supervised learning on the source domain (baseline). Our method, however, stands out with higher Dice scores, lower Hausdorff distances and notably, a higher rate of anatomical validity. Looking specifically at the Hausdorff distance, our method achieves substantially lower scores. This reflects the fact that the output segmentations have smoother borders with less variability. Errors are smaller in cases where the segmentation may be inaccurate. Also, holes and protrusions are almost nonexistent, which is reflected in the anatomical validity scores.



Representative segmentation results from the different methods are presented in Fig.~\ref{fig:segmentation_examples}. Selected images were poorly segmented by the baseline model, therefore examples show the improvement provided by these methods. This confirms the observations from Tab.~\ref{tab:results}. SOTA methods' overall coverage of the valid segmented areas is greater than the baseline, but anatomical inconsistencies remain. RL4Seg conserves anatomically valid shapes while improving segmentation quality.

\begin{figure}[tp] 
\centering
\includegraphics[width=\textwidth]{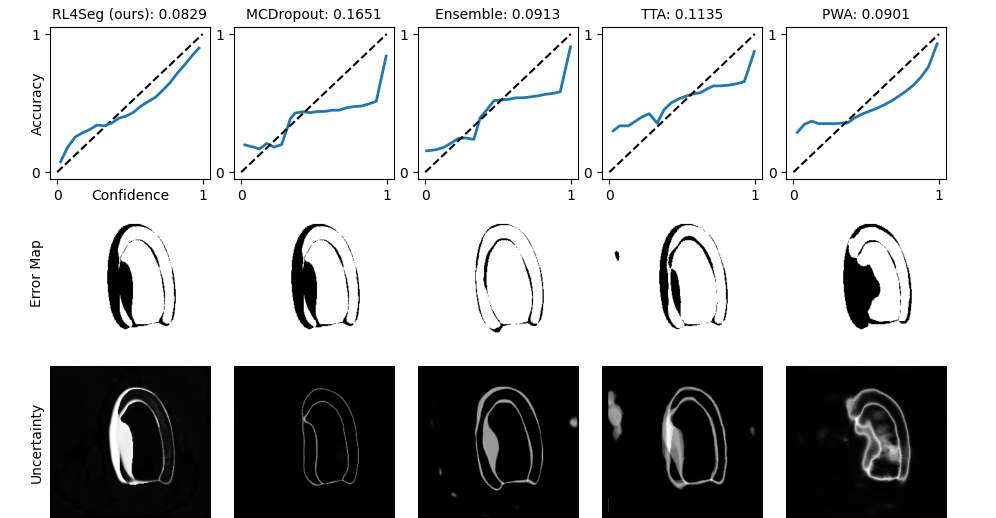} \vspace{-0.4cm}
\caption{Top: Reliability diagrams and ECE for five uncertainty methods, evaluated on the entire target test set. The dashed line represents perfect calibration. Middle and bottom: Examples of error map and corresponding uncertainty map for each method for segmentations from SOTA models on the target test set.\vspace{-0.5cm}} \label{fig:reliability_diag}
\end{figure}

While nnU-Net is also trained exclusively on the source domain, its usage of common voxel spacing and patches may help it generalize better \cite{Isensee2021-nnunet}, thus performing better on the target domain as well. UDAS helps deal with the domain shifts of high-level features present in the images, but not on possible differences of underlying structures between domains. The TS-IT method relies on reliable pixels present in pseudo-labels. Confidence may be underestimated in some regions, leading to holes in the output segmentations. In addition, TS-IT is binary in nature, so the coherence of the overall mask is lacking, as the left-ventricle and myocardium segmentation masks were computed separately.

In general, these methods have reasonable performance on the target domain, but they remain limited by the inconsistencies in the output segmentations. While sufficient data is available for the model to learn anatomical constraints implicitly in the source domain, this is not the case for the target domain. Ignoring the underlying nature of the segmented structure leads to many incoherent segmentations corresponding to images for which the domain shift is the largest. Our method addresses this issue and provides highly consistent outputs, with results approaching intra-expert variability from the source dataset.

\vspace{-0.2cm}
\subsubsection{Uncertainty}
We compared our reward network with two epistemic uncertainty methods, Monte-Carlo Dropout (MCDropout) \cite{gal2016dropout} and model ensembling \cite{NIPS2017_ensembleUNC}, as well as two aleatoric methods,
test-time augmentations (TTA) \cite{wang2019aleatoricTTA} and pixel-wise aleatoric uncertainty (PWA) \cite{NIPS2017_Aleatoric}. Results for uncertainty estimation of predictions on the target domain are presented in Fig.~\ref{fig:reliability_diag}. The best calibrated method is our reward network, with the lowest expected calibration error (ECE)~\cite{ece}. Also, the reliability diagrams~\cite{reliability_diagram} show that our network is the most consistently calibrated through the entire range of output probabilities. 

Perturbations introduced to both the input images and the model in the creation of the reward dataset allow our reward network to model both epistemic and aleatoric uncertainty, therefore performing better. The reward network stands out especially in cases where large errors occur, as many uncertainty methods model uncertainty along the border of the segmented structures.

\vspace{-0.1cm}
\section{Conclusion}
\vspace{-0.1cm}
We have presented RL4Seg, a novel domain adaptation framework using reinforcement learning. It produces both a strong segmentation model, and an accurate uncertainty estimation network without the need for any annotations on the target domain, which perform better than the state-of-the-art. In addition, our method not only limits the number of anatomical inconsistencies in the segmentations while improving metric scores, but also identifies erroneous or uncertain regions in the segmentation masks. 

\paragraph{Acknowledgements.}
\footnotesize The authors acknowledge the support from the Natural Sciences and Engineering Research Council of Canada (NSERC) [551577754, RGPIN-2023-04584] and the Fonds de recherche du Québec en Nature et Technologies [334622], as well as the partial support from the French ANR (LABEX PRIMES [ANR-11-LABX-0063] of Université de Lyon, and the MIC-MAC [ANR-19-CE45-0005] and ORCHID [ANR-22-CE45-0029-01] projects).
\normalsize

%
%
%
\clearpage
\bibliographystyle{splncs04}
\bibliography{bibliography}

\end{document}


\title{Domain Adaptation of Echocardiography Segmentation Via Reinforcement Learning \\ --- \\ Supplementary Materials\vspace{-1cm}}
\titlerunning{RL4Seg: Domain Adaptation for Segmentation via Reinforcement Learning - Supplementary materials}
%
\author{
}
%
%
\institute{
}

\authorrunning{A. Judge et al.}
\maketitle              

\section{Anatomical Metrics}
\vspace{-1cm}
\begin{table}[h!]   
\caption{Anatomical validity rules for segmentations composed of left ventricle (LV), myocardium (MYO) and background (BG) classes. Inspired by \cite{Painchaud_metrics}.}
    \centering
    \small
    \begin{tabular*}{\textwidth}{p{0.4\linewidth} | p{0.6\linewidth}}
    \toprule

    Metric & Description \\

\midrule

Presence of LV & There are pixels of class LV. \\
Presence of MYO & There are pixels of class MYO. \\
LV holes & No holes are present in the LV. \\
MYO holes & No holes are present in the MYO. \\
LV disconnectivity & There is only one LV region. \\
MYO disconnectivity & There is only one MYO region. \\
Holes between LV, MYO & There are no holes between regions of LV and MYO. \\
LV \& BG frontier ratio & Border length between LV and BG is within thresholds. \\
MYO thickness & Ratio between minimal and maximal thickness of the MYO is below threshold. \\
LV width / MYO thickness ratio & Relative width of LV and thickness of MYO walls is between thresholds. \\

    \bottomrule
    \end{tabular*}
    \vspace{-0.6cm}
\label{tab:results}
\end{table}

\section{Ablation Study}

\vspace{-1cm}
\begin{table}[h!]   
\caption{Ablation study covering the creation of the reward dataset ($\mathcal{D}_r$): image transforms (brightness, contrast), weight perturbations (Gaussian noise) and anatomical correction with VAE.}
    \centering
    \small
    \begin{tabular*}{\textwidth}{ccc @{\extracolsep{\fill}} | @{\extracolsep{\fill}} ccc}
    \toprule

    \mcc{\makecell{Image \\ Transforms}} &  \mcc{\makecell{Weight \\ perturbations}} & \mcc{\makecell{Anatomical \\ correction}} & \mcc{Dice (\%) $\uparrow$} &  \mcc{HD (mm) $\downarrow$} & \mcc{\makecell{Anatomical \\ Validity (\%) }$\uparrow$}  \\

\midrule

\checkmark & & & 92.2 & 6.1 & 98.6 \\
& \checkmark & & 91.5 & 6.8 & 90.0 \\
& & \checkmark & 92.1 & 6.3 & 94.8 \\
\checkmark & \checkmark & & 93.0 & 5.8 & 98.2 \\
& \checkmark & \checkmark & 92.7 & 6.0 & 98.4 \\
\checkmark & & \checkmark & 92.1 & 6.0 & \textbf{98.9} \\
\checkmark & \checkmark & \checkmark & \textbf{93.3} & \textbf{5.3} & \textbf{98.9} \\

    \bottomrule  & 
    \end{tabular*}
    \vspace{-0.6cm}
\label{tab:results}
\end{table}

\section{Uncertainty Results}
\vspace{0.6cm}
\begin{figure*}[h!]
\centering
\includegraphics[width=\textwidth]{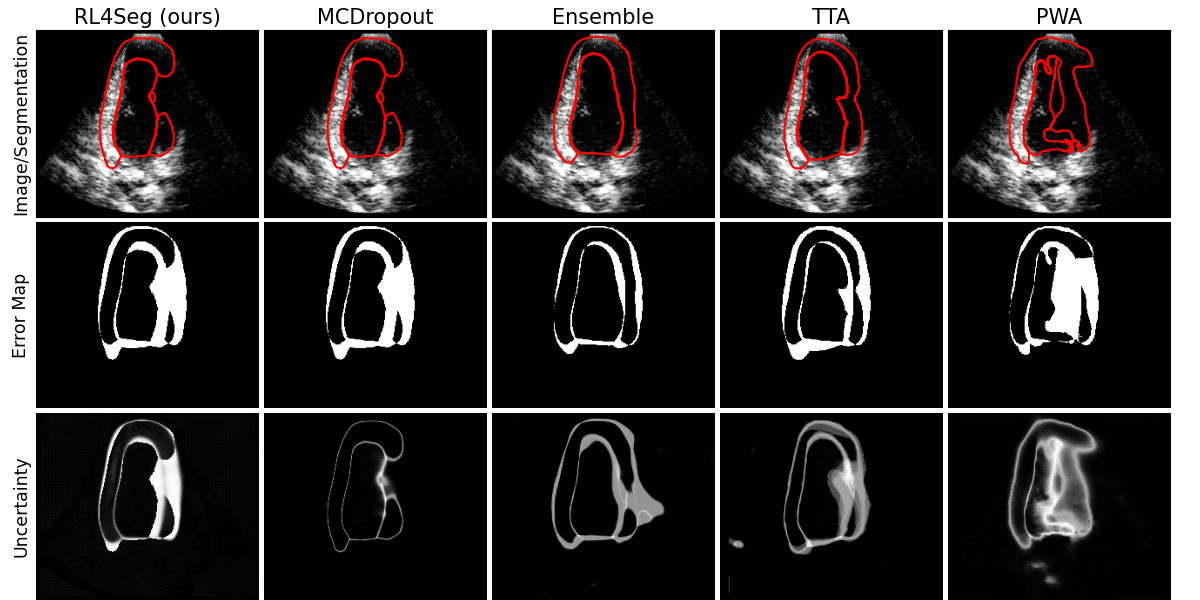}
\includegraphics[width=\textwidth]{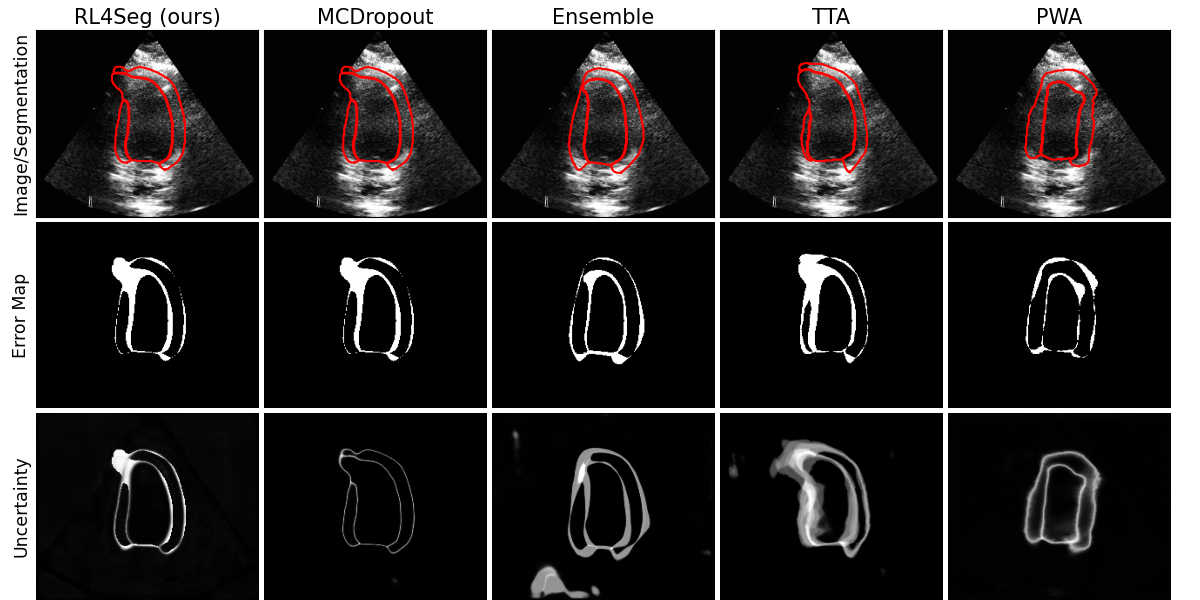}
\caption{Additional uncertainty results for SOTA methods and RL4Seg on different subjects of the target domain. Segmentations with errors (contour presented on image) from baseline models.}
\label{fig:samples}
\end{figure*}

\bibliographystyle{splncs04}
\bibliography{bibliography}